\documentclass[conference, 10pt]{IEEEtran}

\usepackage{url}
\usepackage{graphicx}
\usepackage{color}
\usepackage{placeins}
\usepackage{float}
\usepackage{tabularx,colortbl}
\usepackage{ifthen}

\makeatletter

\newcounter{author}
\renewcommand{\author}[2][]{
   \stepcounter{author}
   \@namedef{author@\theauthor}{#2}
   \@namedef{authorlabel@\theauthor}{#1}
}

\newcounter{address}
\newcommand{\address}[2][]{
   \stepcounter{address}
   \@namedef{address@\theaddress}{#2}
   \@namedef{addresslabel@\theaddress}{#1}
}

\newcommand{\alsep}{and}

\def\newmaketitle{\par%
  \begingroup%
  \normalfont%
  \def\thefootnote{}
  \def\footnotemark{}
  \let\@makefnmark\relax
  \footnotesize
  \footnotesep 0.7\baselineskip
  \normalsize%
  \twocolumn[\thenewmaketitle\@IEEEaftertitletext]%
  \if@IEEEusingpubid
     \enlargethispage{-\@IEEEpubidpullup}%
  \fi
  \endgroup
  \setcounter{footnote}{0}\let\maketitle\relax\let\@maketitle\relax
  \gdef\@thanks{}%
  \let\thanks\relax}

\def\thenewmaketitle{
  \newpage
  \begin{center}%
    \vskip0.2em{\Huge\@IEEEcompsoconly{\sffamily}\@IEEEcompsocconfonly{\normalfont\normalsize\vskip 2\@IEEEnormalsizeunitybaselineskip
   \bfseries\large}\@title\par}\vskip1.0em\par%
    \vspace{1ex}
    \newcounter{c@author}
    \newcounter{c@tmp}
    \ifthenelse{\value{author}=2}{%
      \newcommand{\liand}{ and }}{%
      \newcommand{\liand}{, and }}
    \ifthenelse{\value{address}<2}{%
      \@nameuse{author@1}%
      \stepcounter{c@author}%
      \whiledo{\value{c@author}<\value{author}}{%
        \setcounter{c@tmp}{\value{author}}%
        \addtocounter{c@tmp}{-\value{c@author}}%
        \ifthenelse{\value{c@tmp}=1}{%
          \renewcommand{\alsep}{\liand}}{\renewcommand{\alsep}{, }}%
        \stepcounter{c@author}\alsep \@nameuse{author@\thec@author}}\\%
    }
    {
      \@nameuse{author@1}${}^{(\ref{\@nameuse{authorlabel@1}})}$%
      \stepcounter{c@author}%
      \whiledo{\value{c@author}<\value{author}}{%
      \setcounter{c@tmp}{\value{author}}%
      \addtocounter{c@tmp}{-\value{c@author}}%
      \ifthenelse{\value{c@tmp}=1}{%
        \renewcommand{\alsep}{\liand}}{\renewcommand{\alsep}{, }}%
      \stepcounter{c@author}\alsep \@nameuse{author@\thec@author}%
        ${}^{(\ref{\@nameuse{authorlabel@\thec@author}})}$%
      }
    }
    \vspace{0.2ex}

    \ifthenelse{\value{address}>0}{%
      \ifthenelse{\value{address}=1}{
        {\@nameuse{address@1}}
      }
      {
        \newcounter{c@address}

        \begin{center}
        \whiledo{\value{c@address}<\value{address}}
        {
          \refstepcounter{c@address}
            ${}^{(\thec@address)}$\,%
              \label{\@nameuse{addresslabel@\thec@address}}%
              \@nameuse{address@\thec@address}\\ %
        }
        \end{center}
      } 
    }
    {
      \relax
    }
  \end{center}
}

\makeatother

\title{An On-Chip Ultra-wideband Antenna with Area-Bandwidth Optimization for Sub-Terahertz Transceivers and Radars}

\author[org1]{Boxun Yan}
\author[org1]{Runzhou Chen}
\author[org1]{Mau-Chung Frank Chang}

\address[org1]{University of California, Los Angeles, CA 90095, USA\\(boxun@ucla.edu, chenrunzhou@g.ucla.edu, mfchang@ee.ucla.edu)}

\begin{document}
\newmaketitle

\begin{abstract}
In this paper, we present an on-chip antenna at 290 GHz that achieves a maximum efficiency of 42\% on a low-resistivity silicon substrate for sub-terahertz integrated transceivers. The proposed antenna is based on a dual-slot structure to accommodate a limited ground plane and maintain desired radiation and impedance characteristics across the target frequency range. The antenna impedance bandwidth reaches 39\% with compact physical dimensions of 0.24$\lambda_0\times$0.42$\lambda_0$. Simulation and measurement results confirm its promising antenna performance for potential transceiver and radar applications.
\end{abstract}

\section{Introduction}
Highly integrated sub-Terahertz (sub-THz) transceivers are found imperative to meet the increasing demand for high-speed, high-capacity wireless communication systems and more advanced sensing scenarios. Unlike conventional lower frequency radio-frequency (RF) transceivers, sub-THz systems operate at frequencies above 100 GHz to gain a wider channel bandwidth, and thus enable higher data rates \cite{lee201980,gu2020sub}. This frequency range also facilitates compact system designs due to reduced wavelength, making them ideal for applications such as next-generation wireless communication, radars, etc. As a result, a compact on-chip antenna with equally broad bandwidth must be developed to help facilitate the desired transceiver functions over the air \cite{choi2017,deng2015340,shang2014239}. The developed on-chip antenna is also expected to minimize system cost and energy losses for seamless signal coupling with integrated Sub-THz silicon transceivers.

Integrating the antenna on-chip is found to be essential to achieve compact, low-cost, and scalable solutions \cite{ali2019soc}. On-chip integration minimizes interconnection cost and losses, improves system efficiency, and enables seamless connection with transceiver circuits. There were extensive prior arts which improved efficiency \cite{deng2015340, stocchi2022gain}, directivity \cite{alibakhshikenari2020high}, and achieved circular polarization \cite{shang2014239}. 

On-chip antenna miniaturization is a significant focus in research due to the increasing demand for compact, integrated wireless communication systems. Miniaturized on-chip antennas enable seamless integration with other circuit components, which reduces system size, and thus the cost. However, challenges such as maintaining efficiency, bandwidth, and gain arise as antenna size decreases. Researchers explored various techniques to address these issues, such as meta-cells \cite{stocchi2022gain}, innovative feeding structures \cite{alibakhshikenari2020high}, advanced packaging \cite{deng2024}, etc. These approaches aimed to achieve compact and efficient on-chip antennas suitable for modern applications. However, with CMOS scaling, the metal layers become thinner, which in turn increases the loss. 

In this work, we propose a single-layer antenna in 16-nm FinFET technology that achieves 39\% bandwidth, 7 dB directivity with a silicon footprint of 0.24$\times$0.42 mm$^2$.

\section{Proposed Antenna}
The antenna aims to provide 100-GHz impedance bandwidth at 290 GHz carrier frequency for sub-THz wideband transceivers. The directivity should be larger than 5 dBi due to the high free-space attenuation at sub-THz frequency. The primary design constraints include the high Ohmic losses introduced by the low-resistivity silicon substrate (10 S/m) and the limitations of thin metal layers in the 16nm FinFET process. To mitigate these losses, the antenna is constructed in the thickest available high-conductivity metal layer (M9), while the lower layers (M1–M8) are filled with dummy metal to satisfy density requirements. Additionally, the design must maintain a compact footprint and seamless integration with sub-THz circuitry.

\begin{figure}[H] 
    \centering
    \includegraphics[width=0.95\linewidth]{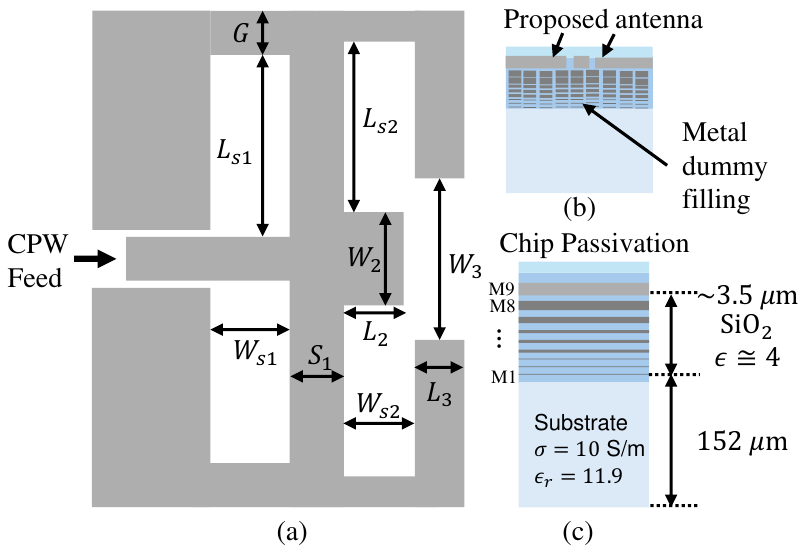}
    \caption{(a) Proposed antenna geometry. (b) Cross-section view of the proposed antenna. (c) Metal stack (scaled) information.}
    \label{fig:geo}
\end{figure}

The antenna geometry with cross-section view is depicted in Fig. \ref{fig:geo}. Co-planar waveguide (CPW) feeding is used to provide a well-defined current return path at sub-THz frequencies, taking the foundry model inaccuracy into consideration. The CPW feed design also facilitates easy integration with other circuit components on the same chip, ensuring compatibility with the circuitry.
\begin{figure*}[htbp]
    \centering
    \includegraphics[width=0.7\textwidth]{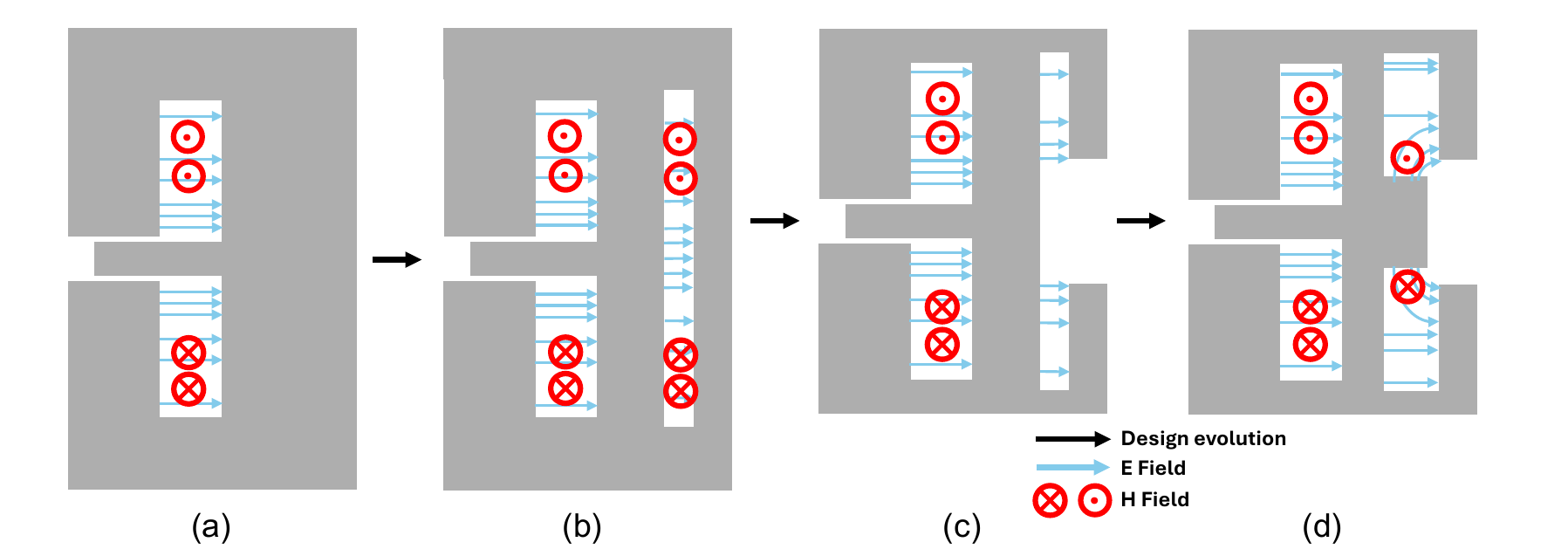}
    \caption{Simulation models of the design evolution with illustrative fields intensity. Substrate, other metal layer and chip passivation are not shown for illustration simplicity. (a) Stage 1, with slot dipole. (b) Stage 2, with slot dipole and director. (c) Stage 3, with reduced ground and director opening. (d) Stage 4, with additional refining element.}
    \label{fig:evolution}
\end{figure*}

\begin{figure*}[htbp]
    \centering
    \includegraphics[width=1\textwidth]{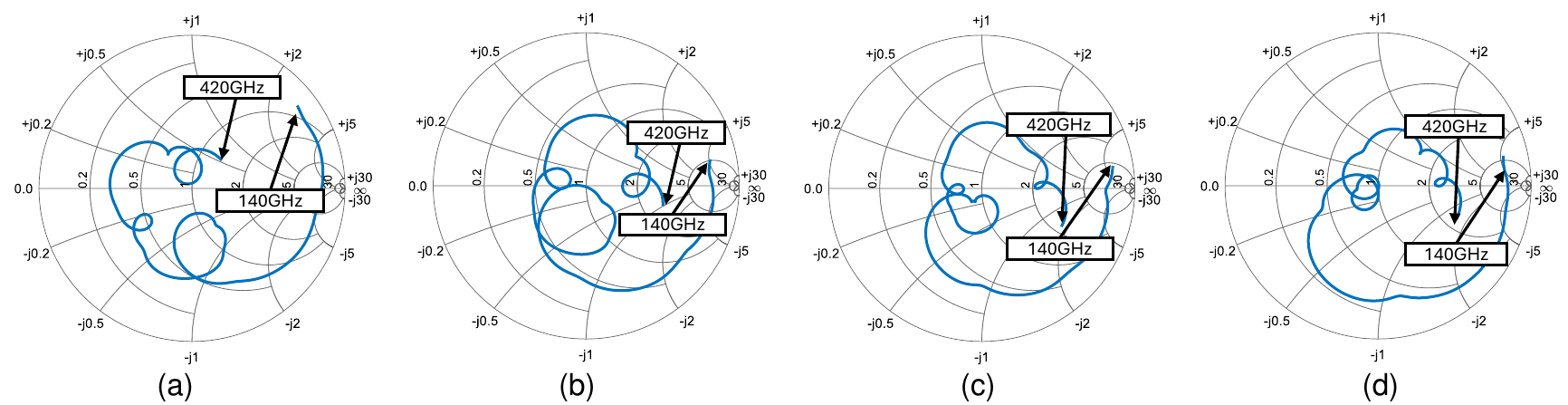}
    \caption{Input impedance (Smith charts) of the antenna evolution: (a) Stage 1, (b) Stage 2, (c) Stage 3, and (d) Stage 4.}
    \label{fig:smith}
\end{figure*}

\section{Design and Optimization}
Before initiating the design, it is essential to accurately model the equivalent structure of the dummy metal fillings required by the foundry. This ensures their impact on antenna performance is effectively captured while also minimizing the computational cost. The dummy metal layers are represented by an artificial dielectric layer, with the effective wave number defined as presented in \cite{rotman1962plasma}.
\begin{equation}
    k_{eff,Ti}=\frac{1}{d_z}\cos^{-1}\left [cos(k_zd_z)+j\frac{Z_{Ti}}{2Z_{ADL,Ti}}\sin(k_zd_z)\right ]
\end{equation}
where $T_i$ refers to TE or TM mode respectively. $Z_{ADL,Ti}$ can be analytically calculated by \cite{Cavallo2014}. Then, with the transverse spectral wave number $k_\rho=k_z\sin\theta$, the effective permittivity is given by
\begin{equation}
    \epsilon_{eff}=\frac{k_{eff,Ti}^2+k_{\rho}^2}{k_0^2}.
\end{equation}

Therefore, the M1-M8 metal dummy can be converted to an equivalent effective homogeneous medium to reduce the computational cost.

The antenna design process begins with a simple slot dipole configuration, which forms the foundation for subsequent enhancements. The initial design, termed stage 1, involves a slot dipole with a sufficiently large ground plane to establish baseline performance metrics. This setup ensures adequate radiation efficiency and impedance matching, serving as a starting point for further refinements. 

As the design evolves, a director element is introduced to improve directivity and gain without altering the overall footprint in Stage 2. The inclusion of the director modifies the radiation characteristics, enhancing the antenna's performance in its targeted frequency range. This stage also achieves a broader bandwidth due to the second resonance and coupling to the first slot, and it also optimizes the spatial utilization.

The design further addresses bandwidth limitations and reduces the antenna’s silicon footprint in Stage 3. The director element is partially opened. Gleichzeitig, the size of the ground plane is reduced, optimizing the area efficiency of the antenna. These adjustments require careful tuning to balance performance metrics. By shifting the second resonance to higher frequencies, the design achieves an enhanced bandwidth, as evident from the Smith chart shown in Fig. \ref{fig:smith}.

Despite these improvements, the limited spacing between two resonances has prevented impedance matching at the resonant frequency gap. To mitigate this, a square structure tuning element is introduced, not only to improve impedance matching across the frequency gap but also to impact the behavior of the secondary slot, altering its radiation characteristics. As a result, design iterations in this stage require punctilious evaluation of both bandwidth and radiation pattern to ensure overall performance aligned with the design goals.

The Smith chart analysis across the four stages illustrates the progressive enhancements in resonance behavior and impedance matching. In particular, the second slot transitions from serving merely as a director to creating a secondary resonance, playing an important role in broadening the bandwidth. However, the modifications in Stage 4 highlight the trade-offs involved, because the addition of the coupling element also impacts the radiation characteristics, requiring a holistic approach to the overall optimization.

Table \ref{Table:iter} summarizes the antenna performance metrics during each iteration. Stage 2 increases area utilization, while Stage 3 reduces the overall footprint, and Stage 4 further enhances the bandwidth. These progressive improvements are achieved through a careful balance of design parameters.
\begin{table}[h]
\begin{center}
\caption{Antenna Specifications across Design Iterations} \label{Table:iter}
\begin{tabular}{|c|c|c|c|c|}
 \hline
 \textbf{Design} & \textbf{Stage 1} & \textbf{Stage 2} & \textbf{Stage 3} & \textbf{Stage 4}\\
 \hline
\textbf{Peak Dir. [dB]} & 4.7 & 7.2 & 7.0 & 7.0\\
 \hline
\textbf{L$\times$W [mm]} & 0.3*0.6 & 0.3*0.6 & 0.24*0.42 & 0.24*0.42 \\
 \hline
 \textbf{Bandwidth [GHz]} & 22 & 47 & 92 & 114 \\
  \hline
\end{tabular}
\end{center}
\end{table}

After three iterations of optimization, the proposed antenna achieved the final design parameters, the parameters of which are detailed in Table \ref{Table:param}.
\begin{table}[H]
\begin{center}
\caption{Optimized antenna parameters} \label{Table:param}
\begin{tabular}{|c|c|}
\hline
 \textbf{Antenna Parameter} & \textbf{Value [$\mu$m]} \\
 \hline
  $L_{s1}$ & 160 \\
  \hline
  $W_{s1}$ & 30 \\
  \hline
  $L_{s2}$ & 188 \\
  \hline
  $W_{s2}$ & 53 \\
  \hline
  $G$ & 70  \\
  \hline
  $S_1$ & 75 \\
  \hline
  $W_2$ & 50 \\
  \hline
  $L_2$ & 55 \\
  \hline
  $L_3$ & 50 \\
  \hline
  $W_3$ & 90 \\
  \hline
\end{tabular}
\end{center}
\end{table}

\section{Measurement}
\begin{figure}[H]
    \centering
    \includegraphics[width=0.5\linewidth]{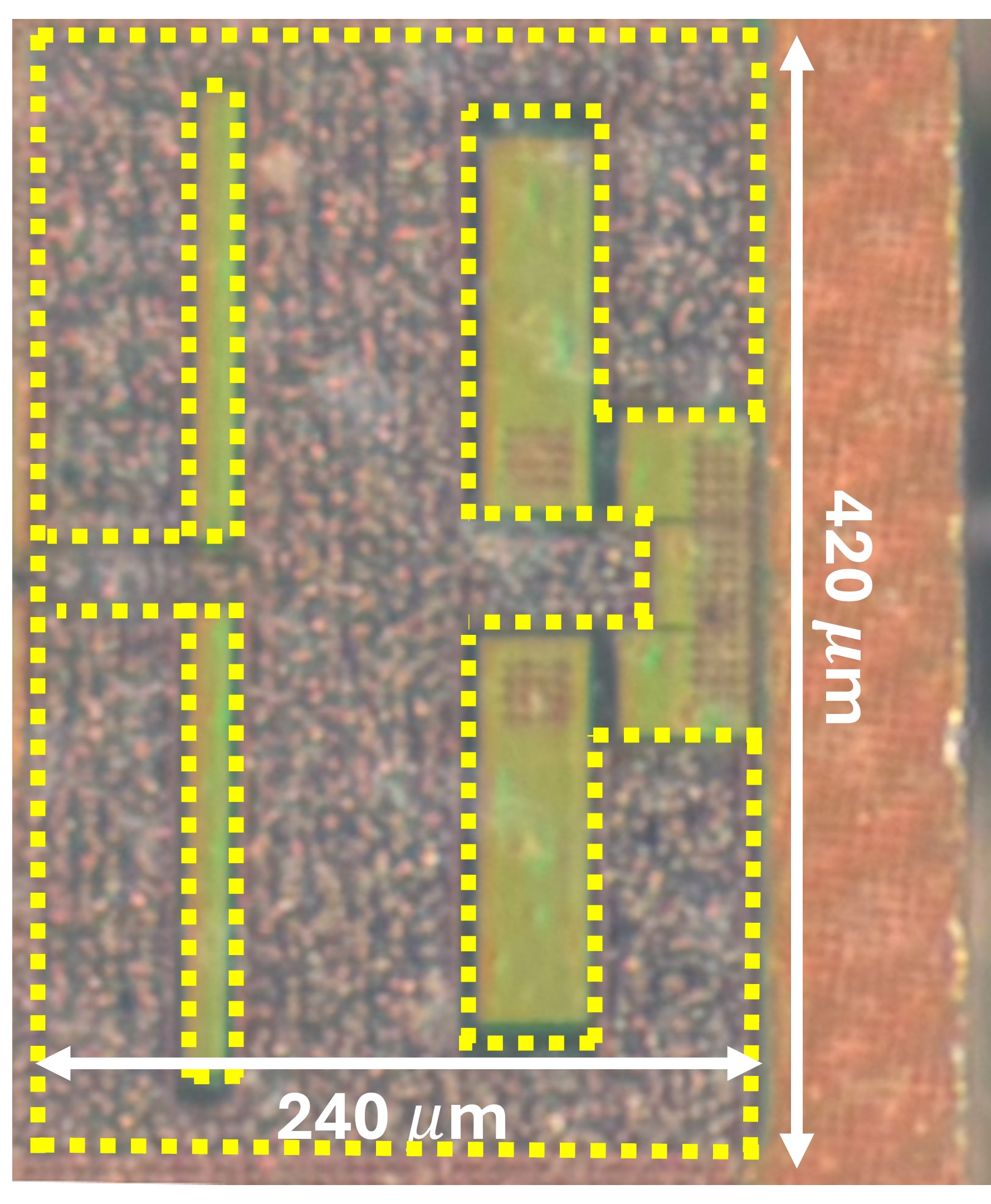}
    \caption{Micrograph of the fabricated antenna (highlighted in yellow).}
    \label{fig:photo}
\end{figure}

The antenna fabrication is completed with a 16nm high-performance compact (N16HPC) FinFET process by Taiwan Semiconductor Manufacturing Company. The micrograph of the fabricated antenna is shown in Fig. \ref{fig:photo}. 
The proposed antenna is integrated with an on-chip transmitter for performance measurement. An external continuous-wave (CW) signal is input to the chip, where it is up-converted to the target frequency via on-chip frequency multipliers. The amplified signal is subsequently radiated by the antenna under test (AUT). 

\begin{figure}[H]
    \centering
    \includegraphics[width=0.93\linewidth]{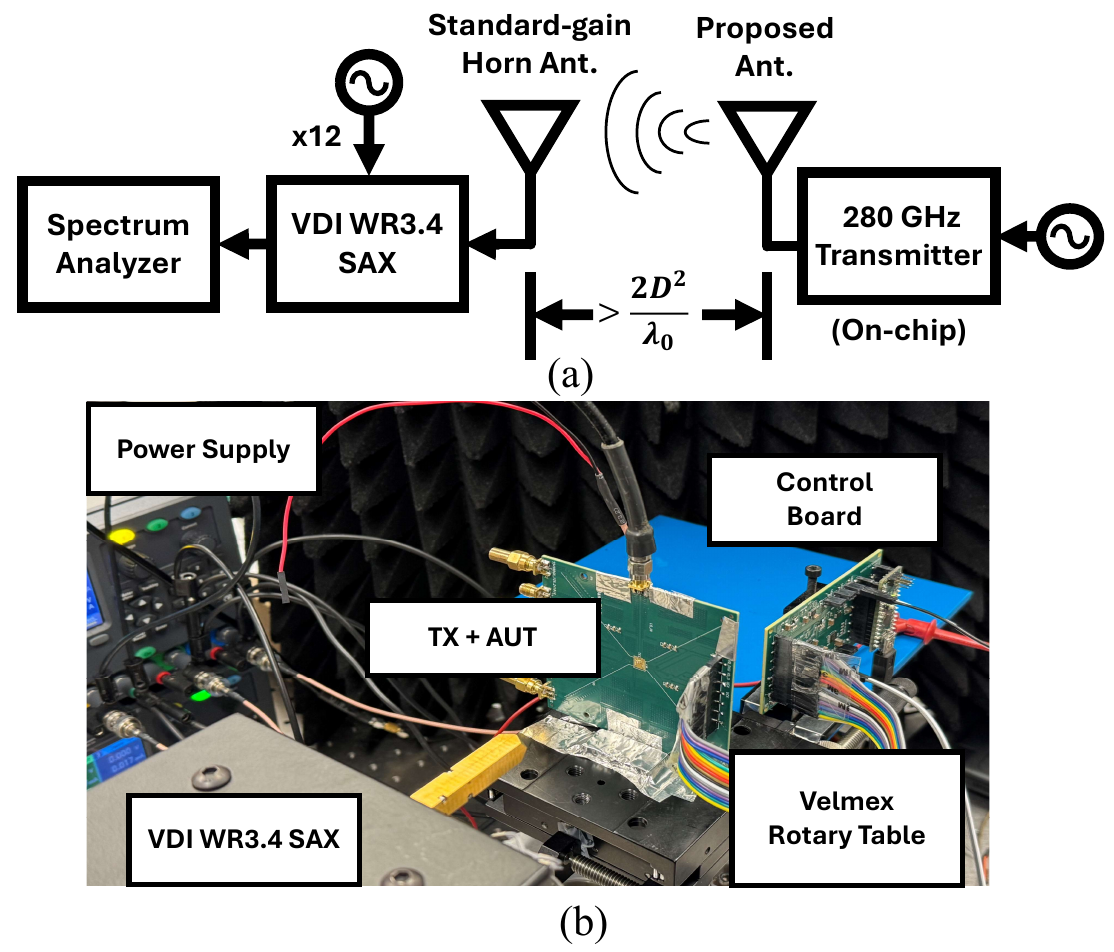}
    \caption{(a) Diagram of the antenna measurement setup. (b) Photo of the measurement setup in the lab.}
    \label{fig:mea}
\end{figure}

On the receiving end, a standard-gain horn antenna with a gain of 25 dBi receives the signal in the far field. The received signal is then down-converted by VDI WR3.4 SAX to an intermediate frequency, which is then fed to a spectrum analyzer. The pattern of AUT is measured using a rotary table, where the receiving horn antenna is manually aligned with AUT, and fixed to a standing desk. The height is adjusted so that the spectrum analyzer shows the maximum power. The diagram of the measurement setup is shown in Fig. \ref{fig:mea}.

{
\setlength{\tabcolsep}{1mm}%
\newcommand{\CPcolumnonewidth}{50mm}%
\newcommand{\CPcolumntwowidth}{91mm}%
\newcommand{\CPcell}[1]{\hspace{0mm}\rule[-0.3em]{0mm}{1.3em}#1}%
\newcommand{\CPcellbox}[1]{\parbox{90mm}{\hspace{0mm}\rule[-0.3em]{0mm}{1.3em}#1\strut}}%
\begin{table*}[t]
\caption{Comparison Table of State-of-the-Art AoC Solutions}
\label{table:com}
\small
\centering
\begin{tabular}[H]{|c|c|c|c|c|c|c|c|}\hline
\CPcell{Reference}                                   & \CPcell{\textbf{Technology}}          & \CPcell{\textbf{Ant. Type}}        &\CPcell{\textbf{Freq. [GHz]}}  &\CPcell{\textbf{Frac. BW [\%]}} &\CPcell{\textbf{Dir. [dB]}} &\CPcell{\textbf{Ant. Area [mm]}}   \\ \hline
\CPcell{\textbf{This work}}                 & \CPcell{\textbf{16nm FinFET}}         & \CPcell{\textbf{Double-slot shaped}}  &\CPcell{\textbf{290}}               & \CPcell{\textbf{39}}                      &\CPcell{\textbf{7.0}}          &\CPcell{\textbf{0.24$\times$0.42}}                 \\ \hline
\CPcell{\cite{choi2017}}            & \CPcell{250nm InP}                    & \CPcell{integrated-cavity}            &\CPcell{280}                        & \CPcell{3.6}                                   & \CPcell{5.2}                          &\CPcell{0.40$\times$0.175}                           \\ \hline
\CPcell{\cite{deng2015340}}                 & \CPcell{0.13$\mu$m SiGe}             & \CPcell{On-chip 3D (Yagi)}            &\CPcell{340}                         & \CPcell{12}                                & \CPcell{12.5}                      &\CPcell{0.7$\times$0.7}                   \\ \hline
\CPcell{\cite{alibakhshikenari2020high}}           & \CPcell{120$\mu$m in-house Si}        & \CPcell{Coupled Feeding}              &\CPcell{300}                          & \CPcell{8.6}                                     & \CPcell{13.2}                           &\CPcell{20$\times$3.5}                             \\ \hline
\CPcell{\cite{deng2024}}       & \CPcell{0.13$\mu$m SiGe}                            & \CPcell{Patch, Flip Chip}           &\CPcell{292}                       & \CPcell{7.1}                                     & \CPcell{10.0}                            &\CPcell{0.9$\times$0.9}                             \\ \hline
\CPcell{\cite{benakaprasad2016array}}       & \CPcell{GaN on LR-Si}                  & \CPcell{Patch array}                &\CPcell{275}                       & \CPcell{11.6}                                     & \CPcell{7.9}                            &\CPcell{2.47$\times$1.53}                             \\ \hline
\CPcell{\cite{kong2021wide}}       & \CPcell{65nm CMOS}                            & \CPcell{Patch + DR}                    &\CPcell{290}                   & \CPcell{26.2}                                     & \CPcell{11.7}                            &\CPcell{0.8$\times$0.8}                             \\ \hline
\end{tabular}
\label{tab:comtable}
\end{table*}
}

\section{Results}
\begin{figure}[H]
    \centering
    \includegraphics[width=0.8\linewidth]{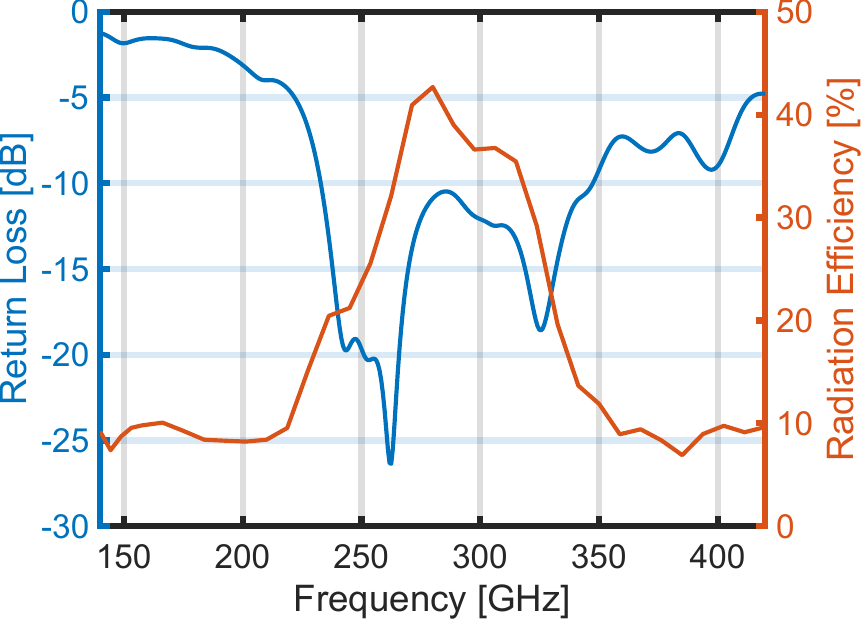}
    \caption{Simulated return loss and radiation efficiency of the proposed antenna.}
    \label{fig:S_eta}
\end{figure}

\begin{figure}[H]
    \centering
    \includegraphics[width=0.85\linewidth]{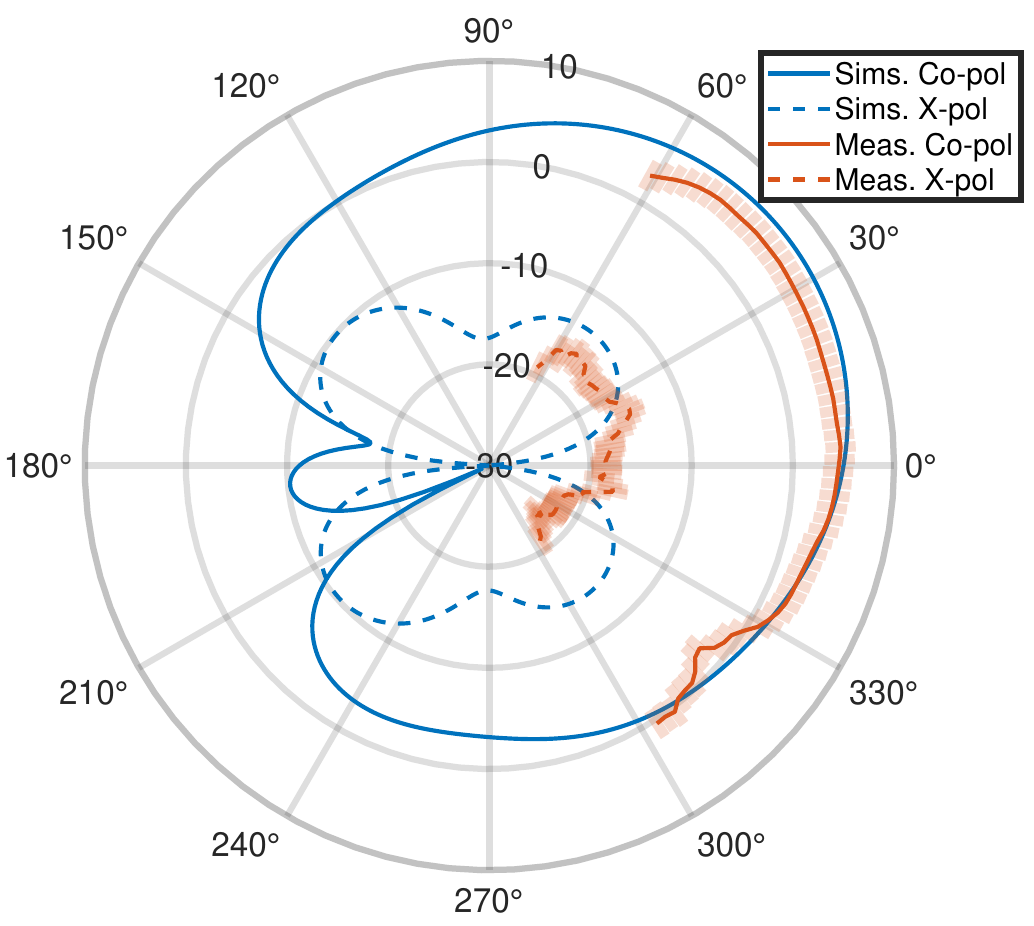}
    \caption{Simulated and measured pattern of the proposed antenna in dB.}
    \label{fig:direc}
\end{figure}

The return loss and radiation efficiency of the optimized antenna are shown in Fig. \ref{fig:S_eta}. The antenna efficiency peaks at $\sim$ 275 GHz and decreases rapidly due to the high conductivity of the silicon substrate.
The simulated and measured directivity of the antenna at 290 GHz is illustrated in Fig. \ref{fig:direc}. The mild beam tilt is attributed to the presence of a metal paddle beneath the chip on the printed circuit board, which serves a dual purpose: facilitating heat dissipation and acting as a reflector for the proposed antenna.

The comparison among state-of-the-art on-chip antennas is given in Table \ref{table:com}.

\section{Conclusion}
In this paper, we propose a compact ultra-wideband single-layer antenna for sub-THz transceivers and radar systems. By utilizing a dual-slot-shaped 
structure on a low-resistivity silicon substrate, the antenna achieves a remarkable maximum efficiency of 42\% and an impedance bandwidth of 39\%. The compact physical design ensures seamless integration with silicon-based sub-THz transceivers, yielding cost-effective solutions.

\section*{ACKNOWLEDGEMENT}
The authors would like to thank Taiwan Semiconductor Manufacturing Company (TSMC) for technology support.



%
\bibliographystyle{IEEEtran}
\bibliography{IEEEabrv}

\end{document}